\documentclass[aps,prl,twocolumn,groupedaddress,floatfix,showpacs]{revtex4-1}
\usepackage{amssymb,amsmath}
\usepackage{graphicx}
\usepackage{subfigure}
\usepackage[english]{babel}
\usepackage{float}
\usepackage{color}

\begin{document}

\title{Experimental observation of inter-orbital coupling}

\author{Diego Guzm\'an-Silva, Gabriel C\'aceres-Aravena, and Rodrigo A. Vicencio}

\address{Departamento de F\'isica and Millenium Institute for Research in Optics - MIRO, Facultad de Ciencias F\'isicas y Matem\'aticas, Universidad de Chile, Chile}
\date{\today}



\begin{abstract}
Inter-orbital coupling refers to the possibility of exciting orbital states by otherwise orthogonal non-interacting modes, a forbidden process in photonic lattices due to an intrinsic propagation constant detuning. In this work, using a femtosecond laser writing technique, we experimentally demonstrate that fundamental and excited orbital states can couple each other when located at different spatial positions. We perform a full characterization of an asymmetric double-well like potential and implement a scan method to effectively map the dynamics along the propagation coordinate. Our fundamental observation constitutes also a direct solution for a spatial mode converter device, which could be located in any position inside a photonic glass chip. By taking advantage of the phase structure of higher-order photonic modes and the effective negative coupling generated, we propose a trimer configuration as a phase beam splitter ($\pi$-BS), which could be of great relevance for multiplexing and interference-based photonic concatenated operations.
\end{abstract}

\maketitle

The way in which atoms assemble and interact with each other determines their potential to form molecules and matter~\cite{r1}. These interactions are characterized by different atom states called orbitals. In order to understand and explore otherwise occulted properties, diverse areas of research have proposed the creation of artificial atoms in different physical contexts~\cite{r2,r3,r4,r5,r6,r7,r8,r9}. This includes numerical and experimental research suggesting new materials and the possibility of observing new exotic interactions, as solutions for transport and localization of energy, key scientific goals in science. However, although theoretical researchers have assumed the basic and fundamental concept of inter-orbital interactions as obvious and trivial~\cite{r10,r11}, no systematic experimental evidence is found in literature. In this Article, we demonstrate, both numerically and experimentally, that orthogonal orbital states, located at different two-dimensional (2D) photonic atoms, couple each other by an evanescent interaction. We show that symmetric S-like wave functions could couple to P-like orbitals and also to higher order states. We use a femtosecond (fs) laser technique~\cite{r12} to fabricate elliptical 2D waveguides having different propagation constants, which is analogous to different orbital energies. We construct double-well photonic asymmetric potentials~\cite{r13}, such that the on-site energies of orthogonal orbitals can be finely tuned and coupling between them becomes possible. Our results offer a new way of studying lattice dynamics where, historically, S-like orbitals have been assumed mostly. An inter-orbital coupling gives the opportunity of tuning the sign of hopping between atoms, increasing the available tools for researchers of different areas and, also, amplifying the possibilities to discover new phenomena~\cite{r14,r15}.

The paraxial wave equation in optics and the Schr\"odinger equation in quantum mechanics have exactly the same mathematical form and, therefore, their mathematical solutions must be equivalent~\cite{r2}. Theory tells us that a potential-well structure will always have a bound state~\cite{r1,r16}, independently of its size (width and depth). However, if we want to excite and observe higher-orbital states, we must satisfy a specific condition that, for one-dimensional (1D) structures, reads as
\begin{equation}
8mV_o a^2 \geqslant h^2\hspace{0.5cm}\text{and}\hspace{0.5cm}8n_o \Delta n\ w^2 \geqslant \lambda^2\ ,
\label{c1}
\end{equation}

\noindent for quantum mechanics and optics, respectively. Here $m$ is the particle mass, $V_o$ and $a$ are the potential depth and width, respectively, and $h$ the Planck constant; while $n_o$ corresponds to the bulk refractive index, $\Delta n$ is the refractive index contrast, $w$ is the waveguide width, and $\lambda$ the wavelength. We notice that both relations are equivalent and depend on phenomenologically similar parameters: if the size of atoms/waveguides is larger than some lower bound, higher orbitals are allowed to exist. Specifically in optics, we notice that a larger index contrast and/or a wider waveguide facilitates the generation of a larger number of eigenstates and waveguides could become multimode depending on the excitation wavelength~\cite{r16,r17}.


Resonance is a key concept in physics and many interactions can be well understood by finding conditions for matching the eigen-frequencies of a given system. Of course, complex resonance problems including different types of interactions~\cite{r18} are neither trivial nor simple to be described, but the fundamental idea continues relying on matching or tuning a set of system parameters. 
Similar to what happens in any physical problem having waves and restrictions, eigenmodes are classified by their symmetry, where the ground state~\cite{r10} is always symmetric with respect to the center of the atom/waveguide. This state is the so-called TE00 mode in optics~\cite{r19} or ground state in quantum mechanics~\cite{r1}, and has a Gaussian-like profile and larger propagation constant $\beta_S$ (lower energy). A second state corresponds to a wavefunction which has a node at the center and, as a consequence, a different phase at different lobes. In optics this state is called TE01 or TE10 depending on its orientation~\cite{r19}, while in atom physics is usually called P-state, having also some variations depending on spatial orientation~\cite{r20}. For simplicity we use the Hydrogen atom-like nomenclature: S and P states for TE00 and TE01/TE10 modes, respectively.

\begin{figure}[htbp]
\centering\includegraphics[width=0.9\columnwidth]{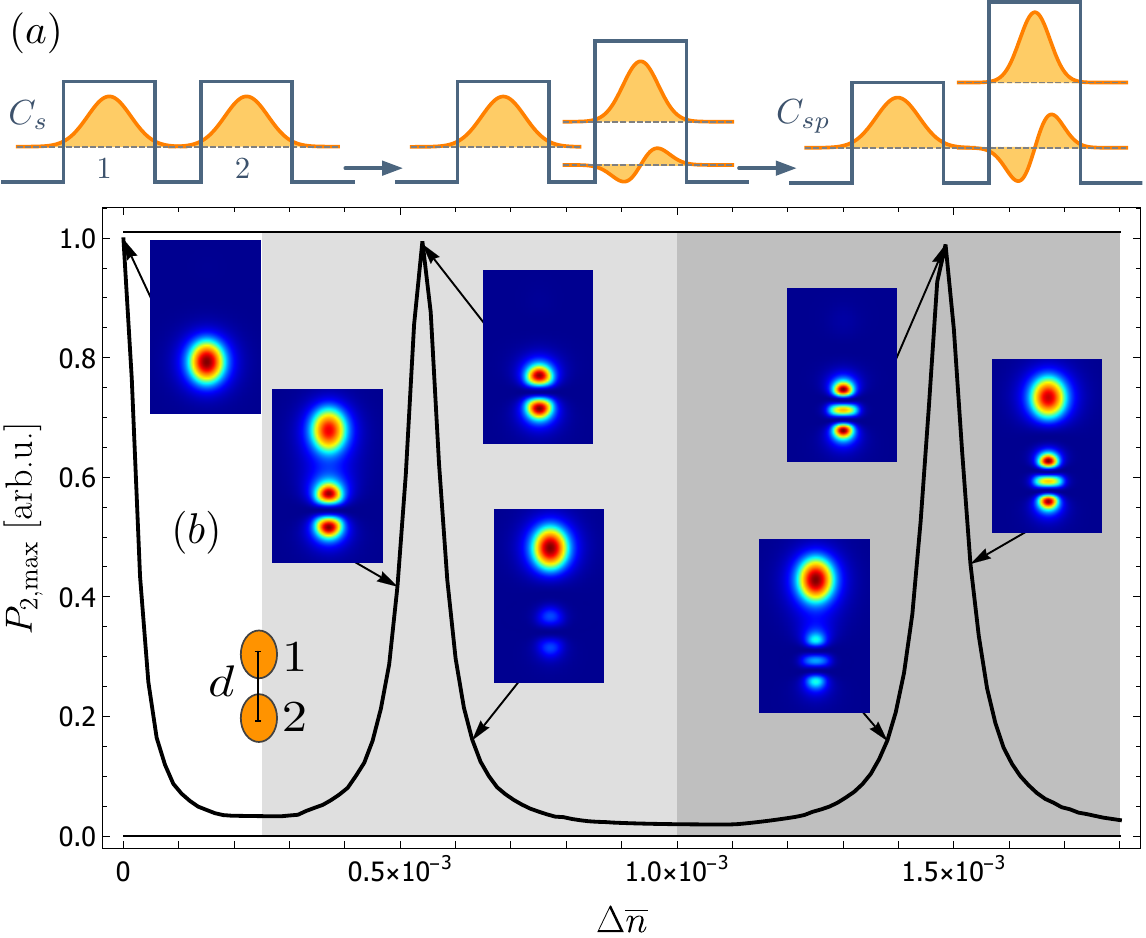}
\caption{a) Propagation constant tuning concept. $C_s$ and $C_{sp}$ correspond to coupling constants in a discrete approach, and define an interaction dynamical scale proportional to $1/C_{s,sp}$. (b) Numerically computed normalized maximum power at waveguide 2 versus refractive index contrast $\Delta \bar{n}$, for two waveguides separated a distance $d=20\ \mu m$. Insets show numerical intensity profiles at indicated regions. White, light gray, and gray shaded areas indicate one, two, or three modes supported at waveguide 2, respectively. A sketch of a vertically oriented dimer configuration is included.}
\label{f1}
\end{figure}

In order to explicitly show inter-orbital coupling, we follow a simple theoretical idea suggested for cold atoms systems~\cite{r13}, as sketched in Fig.~\ref{f1}(a) considering a photonic dimer configuration. Single and multimode waveguides have, in principle, different propagation constants for different orbital states where, for example, $\beta_P<\beta_S$ (in general, propagation constants are always lower for higher-order modes). Therefore, by modifying the local structure of a given waveguide, we can tune its properties and find conditions for an optimal orbital interaction, such that orthogonal states located at different sites have a closer - if not equal - propagation constant. This process would imply an effective dynamical transformation from a S-S interaction into a S-P one, what is the main goal of the present work. We investigate this theoretical concept by, first, numerically integrating a paraxial wave equation considering a coupler (dimer) configuration, which consists of two vertically oriented elliptical waveguides separated by a given distance $d$ (see sketch in Fig.~\ref{f1}(b) and Supplemental Material~\cite{r20p5} for details). We define a relative contrast $\Delta \bar{n}\equiv \Delta n_2-\Delta n_1$, where $\Delta n_1$ and $\Delta n_2$ correspond to refractive index contrasts at waveguides 1 and 2, respectively. When both waveguides are equal, a perfect evanescent coupling occurs between S states~\cite{r2,r12,r21}. In Fig.~\ref{f1}(b) we show that for $\Delta \bar{n}=0$ almost $100\%$ of power is effectively transferred to waveguide 2, as expected considering a coupled-mode (discrete) approach~\cite{r2}, which is governed for coupling constant $C_s$ (see Supplemental Material~\cite{r20p5}). By increasing the relative contrast $\Delta \bar{n}$, an effective detuning is produced in between the S modes at different waveguides. This diminishes the effective interaction between S states at different positions and the transferred power is drastically reduced as Fig.~\ref{f1}(b) shows. We observe a minimum transference for a contrast $\Delta \bar{n}\approx 0.25\times 10^{-3}$. This occurs at the parameter region where the second waveguide starts supporting two modes, as condition (\ref{c1}) suggests. A detuning in propagation constants produces a tendency to localization, due to the creation of a non-symmetric dimer system with corresponding non-symmetric eigenstates. An on-site detuning reduces the effective coupling interaction and the energy tends to remain trapped at the input site. This is phenomenologically similar to a nonlinear dimer~\cite{r22,r23}, where the nonlinear response of the system effectively decouples the waveguides due to an effective refractive index change.

By further increasing the refractive index contrast at site 2, in a region where waveguide 2 already supports two modes, we find that power transfer increases abruptly. We notice that this enhanced transfer is now occurring due to an interaction between S and P states, as inset profiles show in Fig.~\ref{f1}(b). This implies that the effective inter-orbital interaction is switched on, with a coupling constant $C_{sp}$ now governing the dynamics. We find a new peak at $\Delta \bar{n}\approx 0.53\times 10^{-3}$, where the S power at the first waveguide is almost completely transferred to a P mode at waveguide 2. This sharp peak is an indication that an exact condition for a perfect energy transfer is hard to be achieved numerically and, even more, experimentally (peak at $\Delta \bar{n}=0$ is sharp too, but easier to calibrate as both waveguides are equal). Before and after this peak, the transferred energy at site 2 decreases abruptly due to a detuning between S and P propagation constants, as described in Supplemental Material~\cite{r20p5} for a non symmetric dimer model. By further increasing contrast $\Delta \bar{n}$, we observe that the transferred power reduces to a minimum again, this time at $\Delta \bar{n}\approx 1.0\times 10^{-3}$, in a region where the lower waveguide starts supporting a third state as expected from condition (1). By further increasing the relative index contrast, we find a new resonant peak at $\Delta \bar{n}\approx 1.5\times 10^{-3}$. Now, the enhanced transfer of energy at site 2 is due to an inter-orbital coupling between an S state and a third mode, that we simply call ``Tripole'' or ``D'' state (TE20 or TE02 in optics~\cite{r19}). This mode possesses three lobes and two nodes, with the corresponding phase structure and a vertically oriented profile, due to the waveguide ellipticity. Again, the enhanced peak is quite pronounced and the condition for perfect conversion is hard to be achieved numerically, and even more experimentally. So, in principle, this concept could be applied for the excitation - and controlled generation - of any excited state, as soon as the propagation constants at different sites match (see \cite{r20p5} for more details).


\begin{figure}[t!]
\centering\includegraphics[width=0.9\columnwidth]{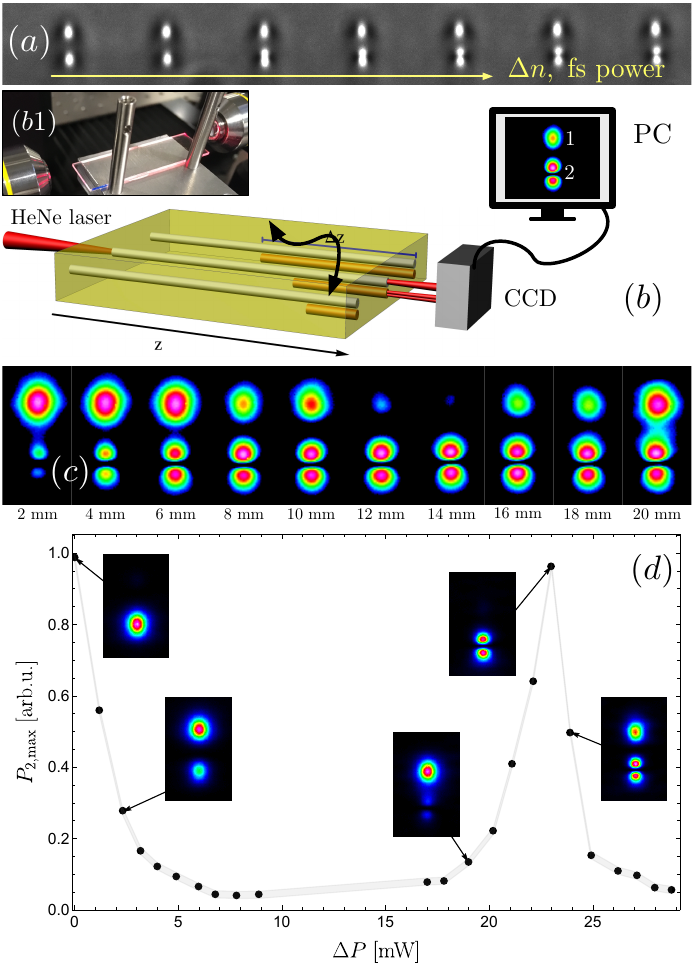}
\caption{(a) Microscope image after white light illumination of dimer examples fabricated using a fixed (variable) writing power for upper (bottom) waveguides. (b) Experimental setup for characterizing waveguides and dimers. P (lower, orange) waveguides have a shorter propagation length $\Delta z$. (b1) Photograph of the experimental setup. (c) $z$-scan showing inter-orbital coupling oscillation at different propagation lengths $\Delta z$ (indicated below each sub-figure). (d) Maximum normalized transferred power at waveguide 2 versus fs average laser power difference ($\Delta P$). Every data point was taken at a different distance}. Insets show experimental output intensity profiles at parameters indicated by arrows.
\label{f2}
\end{figure}


We study optical waveguides as a simil of 2D photonic atoms by using a femtosecond-laser writing technique~\cite{r12} (see Supplemental Material~\cite{r20p5} for fabrication details). We fabricate several photonic dimer configurations as the examples presented in Fig.~\ref{f2}(a), where bottom waveguides become multimode as the writing power increases (see arrow). In order to characterize the asymmetric dimers, we define the following procedure: (i) single-mode waveguides are fabricated using a fixed writing power $P_1=78$ mW along the whole sample ($50$ mm long); (ii) we set a writing power ($P_2$) for waveguide 2 and fabricate 8 dimers for this power, having 8 different final lengths ($\Delta z$) in the interval $\{6,20\}$ mm, with a step of 2 mm, as sketched in Fig.~\ref{f2}(b). From a discrete model, we expect a cosine-like dynamics along the propagation coordinate, which strongly depends on the coupling constant and effective detuning. As a consequence, the maximum transferred power is obtained at different $z$-values and the implementation of a $z$-scan configuration is mandatory in our experiment. We characterize these dimers by focusing a horizontally polarized HeNe laser beam at the input single mode waveguide, as described in Fig.~\ref{f2}(b). In Fig.~\ref{f2}(c) we show an example of the oscillatory (cosine-like) dynamics for $\Delta P\approx 23$ mW. We clearly observe how light from the S (upper) waveguide starts to, first, weakly excite a P state at the bottom waveguide, while for a larger propagation distance the transferred power increases, with a maximum transference at $\Delta z=14$ mm. Fig.~\ref{f2}(c) directly and quite clearly shows the effective transformation from a S mode into a P state. This observation demonstrates experimentally the theoretical concept described in Fig.~\ref{f1} and gives a strong experimental support for the existence of inter-orbital coupling in any physical system having orbital-like states. We clearly observe a dynamical oscillation along $z$ as a proof of a periodic energy transfer mechanism between neighboring coupled waveguides, which is also a strong support for the study of tight-binding (discrete) models including inter-orbital interactions.

In order to construct a complete parameter space, we analyze several intensity profiles (as the examples shown in Fig.~\ref{f2}(d)-insets) for 23 different writing powers $P_2$. Then, we look for the maximum transferred power at waveguide 2, which naturally occurs at different distances~\cite{r20p5}. We integrate the intensity at upper and lower halves of each image and obtain a normalized value. Then, we look for the maximum transference for the 8 different propagation lengths, as a result of the $z$-scan configuration. Fig.~\ref{f2}(d) shows our compiled experimental results for maximum transferred power at waveguide 2 versus the fs average laser power difference, which is defined as $\Delta P\equiv P_2-P_1$. For $\Delta P=0$, we clearly observe that two identical single-mode waveguides have an excellent transference of energy, with almost $100\%$ of efficiency. Then, by continue increasing the writing power $P_2$, we observe how the transferred power decreases abruptly, what is in perfect agreement with direct numerical simulations [see Fig.~\ref{f1}(b)], and with a discrete theoretical approach~\cite{r20p5}.

As our main goal is the observation of inter-orbital coupling, we continue increasing $P_2$ and, theferore, $\Delta \bar{n}$ (in our experiment, $\Delta \bar{n}\sim\Delta P$~\cite{r20p5}). For $\Delta P\approx 7$ mW, we observe that the transferred power starts to increase weakly [see Fig.\ref{f2}(d)], however this increment is in the order of the experimental error. We observe a transition region with less than $10\%$ of transfer, for a large $\Delta P$ range. However, for $\Delta P\geqslant 19$ mW we observe a clear increasing tendency and the excitation of a weak P mode at the second waveguide, as a first indication of inter-orbital interaction. While continuing to increase the writing power $P_2$, we observe an abrupt increment of energy at the bottom waveguide, with a clear peak at $\Delta P=23$ mW, with more than $95\%$ of transference. This peak constitutes a concrete experimental proof for the excitation of an orthogonal state at a neighbour waveguide, based on the theoretical concepts described in the previous section. This observation is a result of transforming a given single-mode waveguide into a multimode one, at a given wavelength. We are indeed observing experimentally a propagation constant tuning process, which allows switching on the effective interaction between otherwise non-interacting orthogonal states. The experimental intensity image at peak [inset profile in Fig.~\ref{f2}(d)] shows a perfect generated dipole at the bottom waveguide. This simple observation constitutes two main outcomes of our work: the demonstration of inter-orbital coupling and a photonic mode converter. The first one is quite a fundamental result: it gives support to different theoretical ideas~\cite{r10,r13,r14} coming from diverse research areas~\cite{r2,r3,r4,r5,r6,r7,r8,r9} and it suggests the possibility of studying new lattice configurations considering hybridized interactions~\cite{r24,r25,r26,r27,r28}. Without the demonstrated tuning mechanism, orthogonal states simply do not interact on a lattice and hybridized physics is simply not possible on a linear regime. The second outcome of our observation is the generation of a simple and concrete method for exciting higher-order spatial states inside a photonic chip, with perfect spatial controllability that can be a key of success for concatenated photonic operations~\cite{r29,r30,r31,r32,r33,r34}.

\begin{figure}[htbp]
\centering\includegraphics[width=0.9\columnwidth]{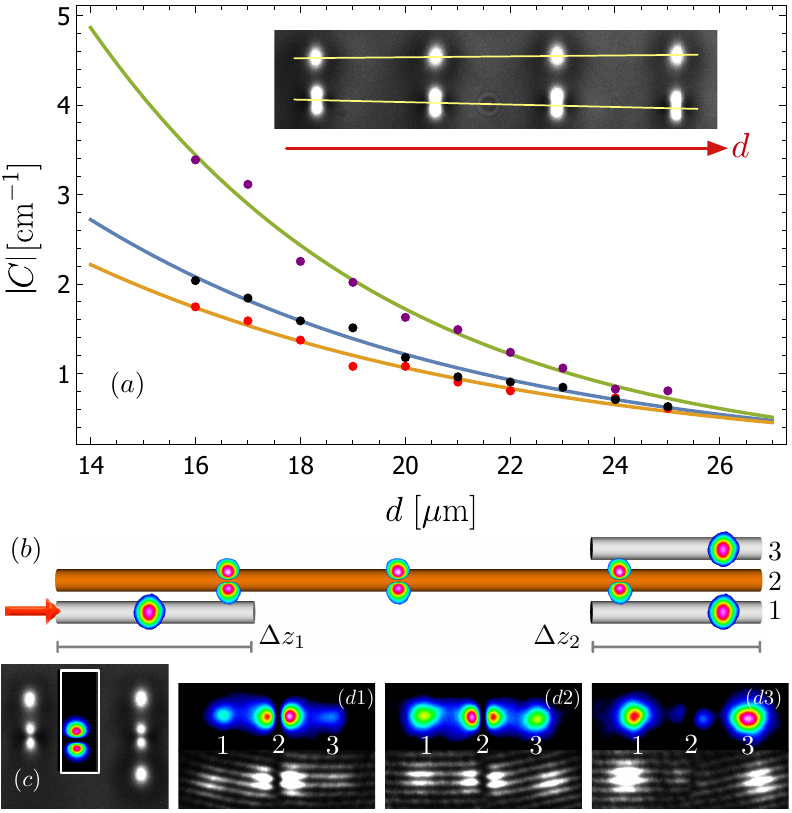}
\caption{(a) Absolute coupling constant $|C|$ versus distance for S-S (red dots), S-P (black dots), and P-P (purple dots) interactions, with lines representing exponential fits. Inset shows a white light microscope image for S-P dimers having different separation distances $d$. (b) SP-dimer + SPS-trimer configuration to demonstrate a phase beam splitter. (c) White light microscope image for a SP-dimer and a SPS-trimer. Inset shows the output profile for the SP dimer only. (d1)--(d3), Output images, for $\Delta z_2=2,4,6$ mm, respectively, after laser excitation of an S waveguide at the input facet. Top and bottom images correspond to intensity and phase profiles, respectively.}
\label{f3}
\end{figure}

Coupled-mode theory~\cite{r2} and tight-binding approximation~\cite{r1} show that modes at different sites/atoms evanescently interact by a coupling constant $C$. In Fig.~\ref{f3}(a) we show an experimental characterization of coupling coefficients for different separation distances and for different orbital states (see examples of SP dimers at Fig.~\ref{f3}(a)-inset and \cite{r20p5} for more details). We observe a clear exponentially decaying tendency~\cite{r12} for all coupling coefficients. As profiles in Fig.~\ref{f2}(c) show, the S mode wavefunction, although being broad, is very well trapped at the waveguide region, while the P mode, in general, occupies a larger vertical area. Therefore, we expect that $|C_s|<|C_{sp}|<|C_p|$, as our experimental data shows in Fig.~\ref{f3}(a) for any distance $d$ (in general, $C_s>0$ and $C_p<0$, while $C_{sp}$ sign depends on orbital orientation~\cite{r15,r17,r27,r28}). This is the first time that this tendency and dependence are clearly and directly shown experimentally for S and P states, giving a strong support for theoretical studies considering different and more complex spatial configurations. It is worth mentioning that the tendency shown in Fig.~\ref{f3}(a) strongly depends on waveguide orientation, which in our case is vertical (see Fig.~\ref{f1}(b)-inset).

As an application of the inter-orbital coupling, we experimentally show the different phases induced due to hybridized interactions. As it has been suggested theoretically~\cite{r13,r28}, an inter-orbital interaction produces the appearance of negative coupling constants, which could be of great impact on the study of topology and effective magnetic fields on lattices~\cite{r15,r35}. Therefore, our method looks as a promising and simple way of generating synthetic magnetic fields~\cite{r36} on lattices, which could have an important impact when considering topological properties~\cite{r37} as well. In order to show an experimental evidence for this effect, we fabricate a setup consisting of a SP-dimer plus a SPS-trimer configuration, as sketched in Fig.~\ref{f3}(b). We focus a HeNe laser beam at the single-mode (white) waveguide and generate a perfect P state at the multimode (orange) waveguide, after a coupling distance of $\Delta z_1=14$ mm (Fig.\ref{f3}(c)-left shows an SP-dimer). Afterwards, the P state propagates freely along the $z$ direction for about $30$ mm. Then it acts as an input excitation for a SPS-trimer configuration (see Fig.~\ref{f3}(c)-right) and interacts symmetrically with both neighbouring single-mode waveguides. As expected, inter-orbital coupling occurs again and new S modes are generated back in waveguides 1 and 3, as shown in Figs.~\ref{f3}(d)-top. Interestingly, by taking advantage of the phase structure of the P state we are able to induce a $\pi$ phase shift in between the two generated S states. Figs.~\ref{f3}(d)-bottom show different interferograms (see Supplemental Material~\cite{r20p5} for details) with a clear $\pi$ phase shift in between the right and left parts of the output profiles. This clearly indicates that the coupling interaction to the right and to the left single-mode waveguides are effectively opposite in sign, which is a direct proof for a negative coupling constant~\cite{r15}. This system is also an example for a phase $\pi$ beam splitter ($\pi$-BS), which could be quite useful for interferometric quantum optics~\cite{r33,r39,r40}, considering concatenated operations~\cite{r29}. The output profile also coincides with the flat-band mode of a rhombic lattice~\cite{r41}, which is an important subject of research nowadays in photonic lattices.

In conclusion, we have demonstrated numerically and experimentally an inter-orbital coupling interaction between S and P states on a photonic platform. We first corroborated a resonance picture by numerically integrating a paraxial wave equation and clearly observed the coupling between different orbitals, as a consequence of asymmetrizing a double-well like potential photonic structure~\cite{r13}. This process was implemented with high precision using a fs laser writing technique, allowing the excitation of higher order states and experimentally probing negative coupling interactions. Our method also offers a new technique for exciting higher-order spatial states inside an optical chip, which could have a great impact on multiplexing applications~\cite{r42,r43}. The possibility of locating the mode conversion system in any arbitrary position is a direct advantage compared to lithographic-like techniques~\cite{r39,r44,r45,r46,r47,r48}. Our observation constitutes a fundamental validation of inter-orbital coupling, which could be an important new ingredient for further studies in lattice science; but, also, a key tool for concatenated or multiplexing photonic operations in the classical and quantum regimes.

Authors acknowledge stimulating discussions with Magnus Johansson, Luis Foa and Bastian Real. This work was supported in part by Millennium Science Initiative Program ICN17$_-$012 and FONDECYT Grants 1191205 and 3190601.

{}

\end{document}


\title{Supplementary Material for ``Experimental observation of inter-orbital coupling: supplemental document''}

\author{Diego Guzm\'an-Silva, Gabriel C\'aceres-Aravena, and Rodrigo A. Vicencio}

\address{Departamento de F\'isica and Millenium Institute for Research in Optics - MIRO, Facultad de Ciencias F\'isicas y Matem\'aticas, Universidad de Chile, Chile}
\date{\today}

\maketitle

\section{Numerical modeling}

\begin{figure}[htbp]
\centering\includegraphics[width=11cm]{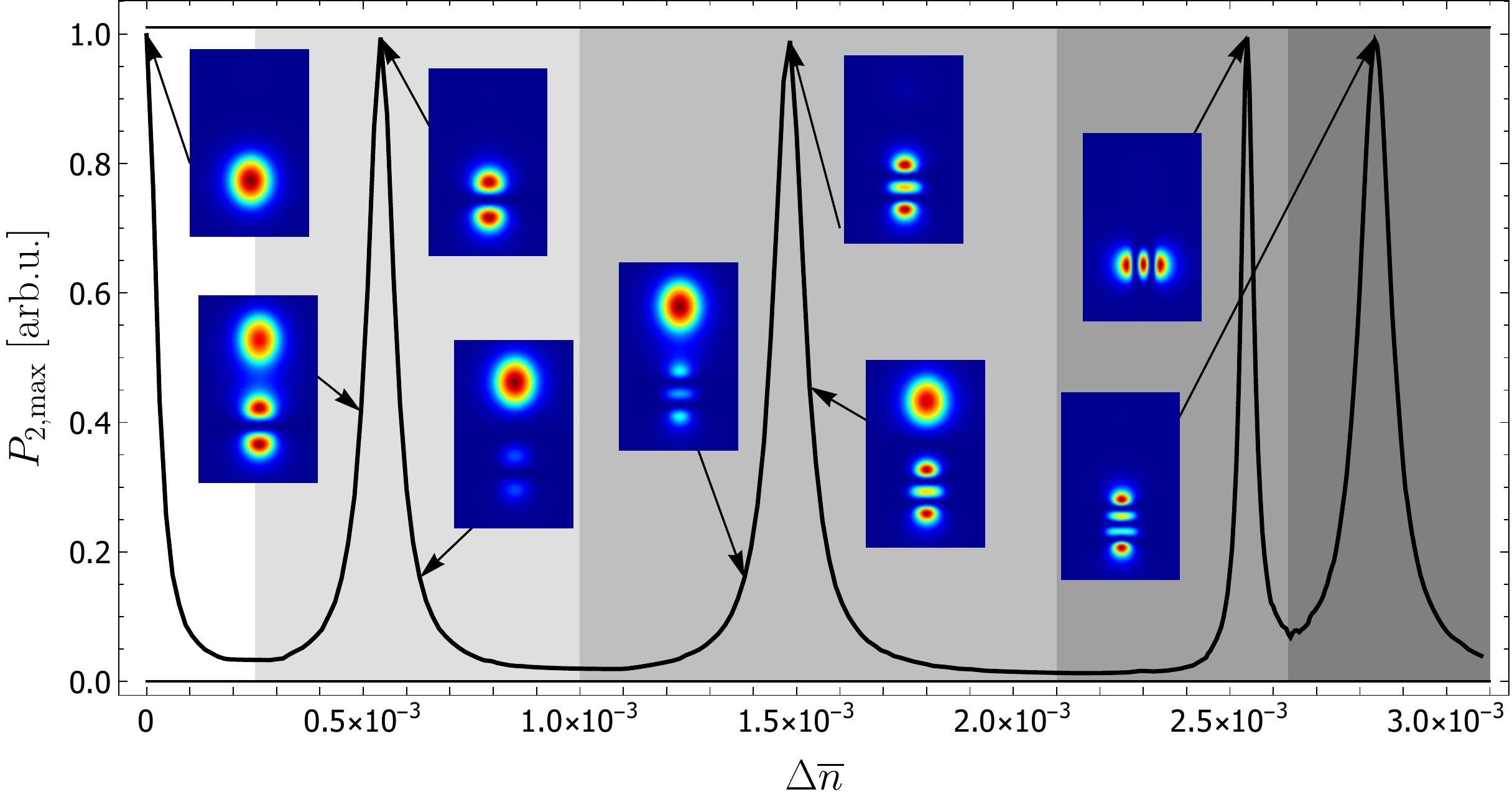}
\caption{Numerically computed normalized maximum power at waveguide 2 versus refractive index contrast $\Delta \bar{n}$, for two waveguides separated a distance $d=20\mu m$. Insets show numerical intensity profiles at indicated by arrows. Different shaded areas indicate one, two, three, four and five modes supported at waveguide 2, from left to the right.}
\label{sf0}
\end{figure}

\noindent
Light propagation through an optical media is well described by a paraxial wave equation of the form, 
\begin{equation}
-i \frac{\partial}{\partial z}\Psi(x,y,z)=\frac{1}{2k_o n_o}\nabla^2_{\bot}\Psi(x,y,z)+k_o\Delta n(x,y)\Psi(x,y,z)\ ,
\label{se1}
\end{equation}
where $z$ is the coordinate along the propagation coordinate and $\Psi(x,y,z)$ is the electric field amplitude (wave-function). $x$ and $y$ the transversal coordinates, $k_o = 2\pi/\lambda$ the wave number in free space, $\lambda$ the wavelength, $n_o$ the refractive index of bulk material. $\nabla^2_{\bot}=\partial_x^2+\partial_y^2$ corresponds to the transversal Laplacian operator, while $\Delta n(x,y)$ defines the transversal refractive index structure inscribed inside the material. Numerical integration of this equation is implemented by means of a beam propagation method (BPM)~\cite{sr1,sr2}, where we numerically simulate the propagation of a given initial profile through a given optical media. The method is initialized by defining input conditions and some optical parameters as wavelength $\lambda = 632$ nm and bulk refractive index $n_o = 1.48$. A vertical distance, center to center, of $d = 20\ \mu$m separates waveguides, and every waveguide has an elliptical profile: $\Delta n_i\tanh[1/(\exp (x^2)+ \exp (y^2))]$, with $i=1,2$. In order to simulate a dimer configuration (as shown in Fig.~1 of main text) the refractive index contrast at waveguide 1 was defined as $\Delta n_1=4.0\times 10^{-4}$ while the waveguide 2 value ($\Delta n_2$) was continuously varied to characterize an asymmetric dimer configuration. We obtain several intensity profiles at different propagation distances, allowing us to track the maximum transferred power at waveguide 2, which naturally occurs at different $z$-values. Following this procedure, we obtain the relation between the maximum intensity value transferred to waveguide 2 versus the change of refractive index contrast at the same waveguide, as shown in Fig.~1(b) (main text). Additionally, we performed extended simulations in terms of $\Delta \bar{n}\equiv \Delta n_2-\Delta n_1$ to evidence the possibility of exciting even higher states. Fig.~\ref{sf0} shows a larger diagram where the waveguide ellipticity is quite evident and where, beside the previously informed excited states, horizontal tripole and vertical quadrupole configurations are also excited. This numerical observation shows quite clearly the tuning concept described in Fig.1(a) (main text) and its potential to increase the lattice complexity in future research.
%
\vspace{-0.25cm}

\section{Asymmetric dimer system}

\begin{figure}[t!]
\centering\includegraphics[width=13cm]{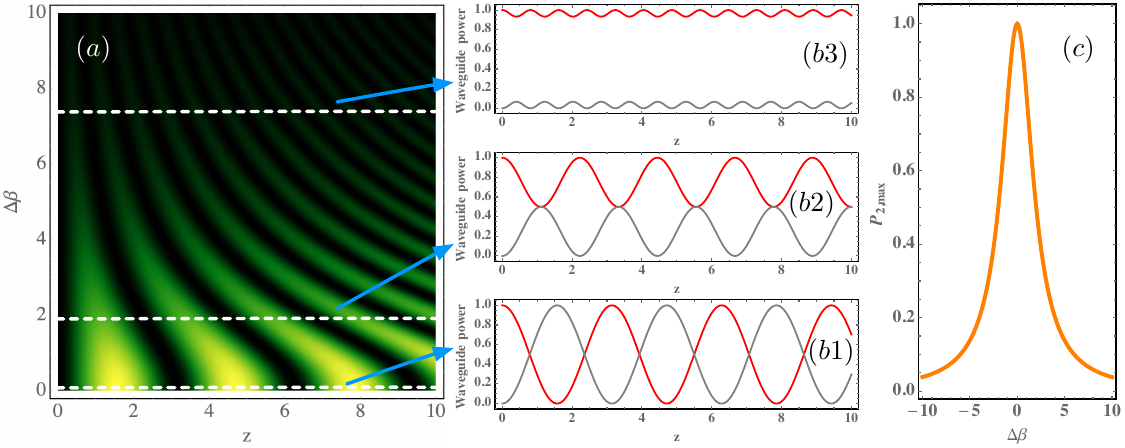}
\caption{(a) Power variation at waveguide 2 versus propagation distance $z$ and detuning $\Delta\beta$. (b) Normalized power at waveguides 1 (red line) and 2 (grey line) for $\Delta\beta=0,2,7.5$, as indicated by arrows and dashed lines in (a). (c) Maximum transferred power at waveguide 2 versus $\Delta\beta$. $C = 1$.}
\label{sf1}
\end{figure}

\noindent
Light propagation in a system composed of two waveguides can be studied numerically using a BPM code (as described in previous section), but there are also approximate methods that capture the main phenomenological features of the system. A broadly used approach is based on coupled mode theory (CMT) or so-called discrete (tight-binding) model. There, a mode at a given waveguide evanescently interacts with nearest neighbour waveguides only, which also present well-defined modes. The dynamics of a dimer system is well described by
\begin{equation}
i \frac{\partial u_1}{\partial z}+\beta_1 u_1 +Cu_2=0\ ,\ \ \ \ i \frac{\partial u_2}{\partial z}+\beta_2 u_2 +Cu_1=0\ ,
\label{se2}
\end{equation}
where $u_n$ and $\beta_n$ describe the electric field mode amplitude and effective propagation constant at the $n$-th site, respectively. $C$ defines a generic coupling constant, which depends on the effective interacting modes and waveguides separation. We define a detuning coefficient as $\Delta\beta\equiv\beta_2-\beta_1$ to characterize the dynamics. Figs.~\ref{sf1}(a)-(c) show that for two identical waveguides ($\Delta\beta=0$) and coupling constant $C_s$, a complete and periodic transfer of energy is observed. 
On the other hand, for $\Delta\beta\neq 0$, a non-perfect transfer of energy (but periodic) is observed due to a mismatch in propagation constants, see examples in Fig.~\ref{sf1}(b) and compiled information in Fig.~\ref{sf1}(c). As long as $|\Delta\beta|$ increases, a lower amount of power is transferred to waveguide 2, which agrees perfectly with the first region shown in Fig.~1(c) (main text), where both waveguides support a fundamental S mode only. Therefore, at this first regime, CMT is perfectly valid as long as waveguides support only one mode. Then, in order to describe an inter-orbital interaction, we simply change the coupling constant from $C_s$ to $C_{sp}$, where $C_{sp}$ describes a coupling between fundamental (S) and first excited (P) states. Of course, this will be valid for a parameter region at which waveguide 2 already supports two modes, as described in Fig.~1(c) (main text). It is important to notice that the system under consideration consists of two waveguides placed vertically. Horizontal S-P coupling is completely zero due to the parity of wavefunctions and perfect cancellation of the superposition integral~\cite{sr3}.
\vspace{-0.25cm}

\section{Femtosecond waveguide writing}

\begin{figure}[htbp]
\centering\includegraphics[width=13.8cm]{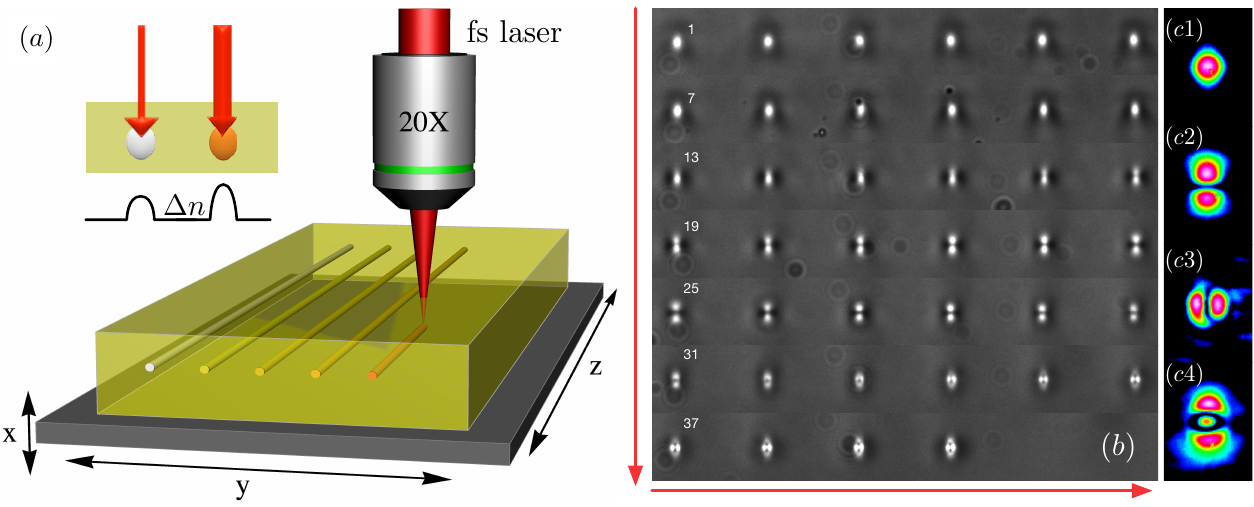}
\caption{(a) A femtosecond laser is focused by a $20\times$ microscope objective inside a borosilicate glass sample. Depending on the laser power, single-mode or multimode waveguides are fabricated inside the material, while a micrometer automatized stage is translated in $XYZ$ directions. A different waveguide color represents a different index contrast. (b) Microscope image at the output facet after white light illumination of 40 individual waveguides. Laser power increases to the right and down, as indicated by red arrows. (c1)-(c4) S, P-vertical, P-horizontal, and D waveguide modes after laser (HeNe) excitation of waveguides 11, 24, 35 and 38, respectively.}
\label{sf2}
\end{figure}

Waveguides were fabricated in borosilicate (Eagle XG) glass wafers of rectangular shape: $50$ mm $\times 10$ mm $\times 1$ mm, using a femtosecond laser writing technique~\cite{sr4,sr5}. To implement this technique, we used circularly polarized ultra-short pulses from a Yb-doped fiber laser (Menlo Systems, Bluecut), operating at a wavelength of 1030 nm, with a pulse width of 230 fs, and a repetition rate of 500 kHz. These pulses were focused into glass wafers using a $20\times$ microscope objective and generating a local increment of the refractive index. Then, by continuously translating the sample (using a $XYZ$ motorized Thorlabs stage), waveguides were written along the whole sample in a very well-defined position. During the fabrication process, the writing velocity was fixed to $v = 0.4$ mm/s. We used a multi-scan technique, where a single waveguide was fabricated by passing 25 times over the same region, where each step is spatially shifted $0.2\ \mu$m in the horizontal direction [Y in Fig.~\ref{sf2}(a)]. The pulse energy was varied from $0.14\ \mu$J ($70$ mW) to $0.24\ \mu$J ($120$ mW).
%
%
Naturally, a larger writing power generates a larger index contrast and, therefore, waveguides are able to support higher-order states, as it was theoretically described in Fig.~1(b) (main text). We start at $70$ mW average fs power and fabricate single mode waveguides at $632.8$ nm (HeNe laser beam). Then, we increase the writing power in steps of $\sim 1.25$ mW up to an average value of $\sim120$ mW, fabricating a total of 40 different waveguides. Using a microscope objective and white light illumination, we image the wafer output facet and obtain the images shown in Fig.~\ref{sf2}(b). We clearly see how waveguides change while increasing the fabrication power and how they start to become multimode. For example, waveguide $17$ already support a second order (white light) P state. This state is clearly observed up to waveguide $\sim 31$. Then, other spatial profiles appear, having horizontal lobes and a mixture of more complex spatial structures. It is important to notice that white light illumination only gives us an indication for a global refractive index change, as this incoherent light source possesses - in principle - all visible wavelengths. Therefore, white light photonic structures observed in Fig.~\ref{sf2}(b) correspond to a mixture between different spatial modes at different colors, where multimode waveguides are first possible for shorter wavelengths [following condition (1) of main text]. Fig.~\ref{sf2}(c) shows S, P and D mode profiles, observed after exciting waveguides with a HeNe laser beam. Higher-order modes are clearly observed with zero intensity in between lobes, as an indication of a staggered phase structure. This experiment clearly shows the excitation of higher-order states as a result of an increasing refractive index contrast.
\vspace{-0.25cm}

\section{Refractive index characterization}

\begin{figure}[htbp]
\centering\includegraphics[width=17.8cm]{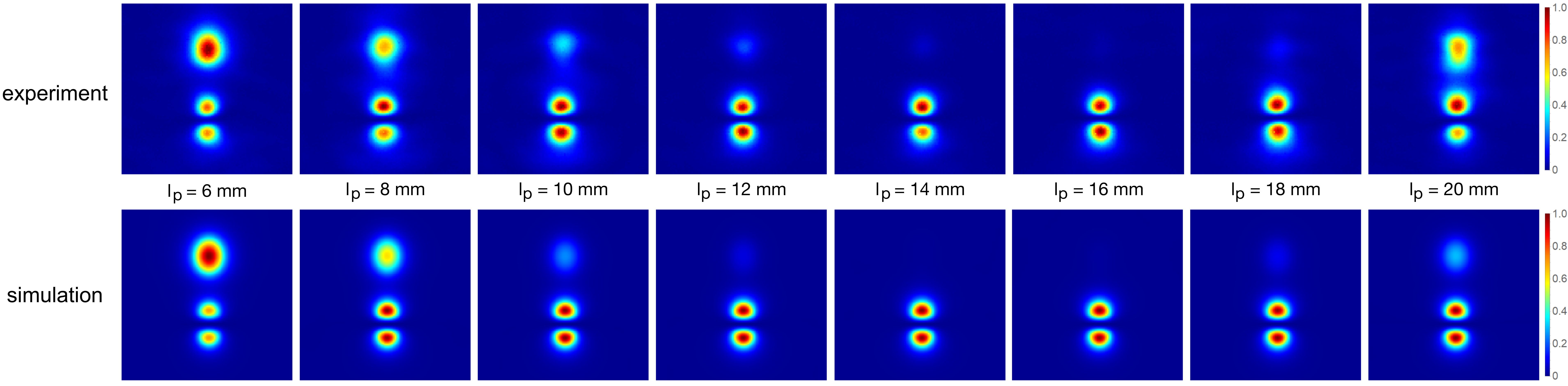}
\caption{Experimental (upper row) and simulated (lower row) intensity output profiles at eigtht propagation lengths $l_p$ (as indicated in each {column}). In this example, top and bottom waveguides are fabricated at a writing power of $78$ and  $101$ mW, respectively. In simulations, we set $\Delta n_1= 4.0\times 10^{-4}$ for waveguide 1 (top) and $\Delta n_2= 9.37\times 10^{-4}$ for waveguide 2 (bottom).}
\label{sf3}
\end{figure}

The number of modes supported by a waveguide depends on its refractive index contrast and width. By increasing the energy of writing pulses, a higher energy is deposited on the sample and a larger refractive index contrast is obtained, as described in Fig.~\ref{sf2}. Then, we proceed to fabricate several vertically oriented dimers, as described in Fig.~2 (main text), where waveguide 2 was fabricated considering different output propagation lengths from 6 mm to 20 mm, in steps of 2 mm. Then, samples were excited with a 632.8 nm focused laser beam and several intensity images were obtained for each writing power. These images are quite useful to characterize the refractive index change at every writing power, which is performed by a direct comparison with numerically obtained profiles (using the BPM described above). Fig.~\ref{sf3} shows an example of this, where an excellent agreement between experimental images and simulated output profiles is observed, at different propagation lengths $l_p$. By doing this for all data presented in Fig.~2(c) (main text), it was possible to find a relation between the experimental laser power difference $\Delta P$ and the estimated refractive index change $\Delta n_2$ at waveguide 2 ($\Delta n_1$ is already fixed as described in Fig.~S3). 
%

\begin{figure}[htbp]
\centering\includegraphics[width=10cm]{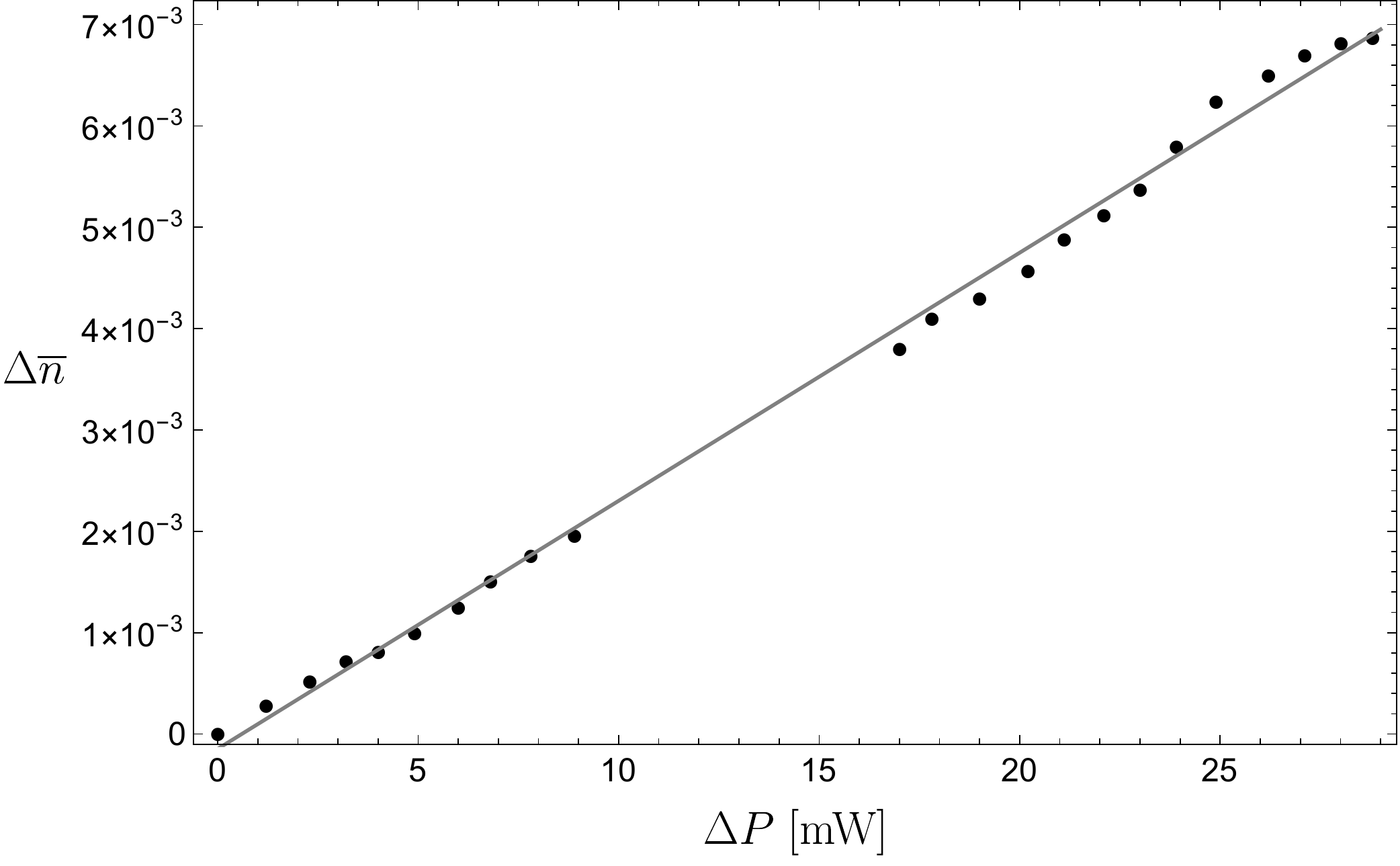}
\caption{$\Delta \bar{n}$ versus $\Delta P$, for the same experimental data shown in Fig.3(c) (main text). A straight line is shown as a fit.}
\label{sf4}
\end{figure}

\noindent
Our characterization is shown in Fig.~\ref{sf4}, where we have added a straight line $\Delta \bar{n}=\alpha\Delta P-\beta$ as a fit, with $\{\alpha,\beta\}=\{2.45,1.47\}\times 10^{-5}$. We observe a clear linear tendency for the power regime studied along this work.

\section{Coupling constant characterization}

\begin{figure}[htbp]
\centering\includegraphics[width=17.8cm]{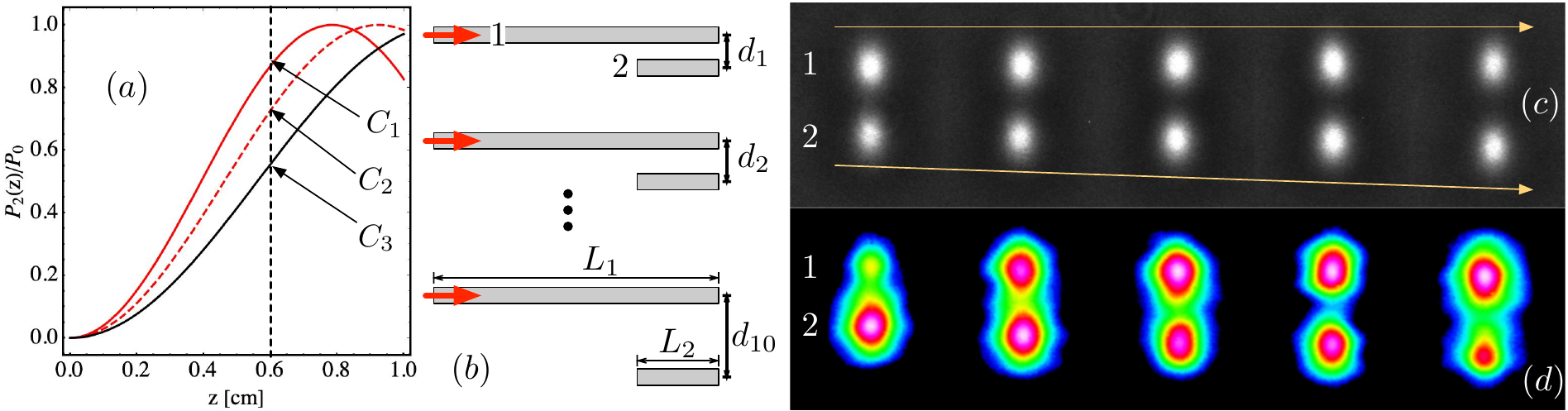}
\caption{(a) Theoretical normalized power at waveguide $2$ versus propagation distance $z$, for three different coupling constants as examples. The vertical dashed line indicates the propagation distance for waveguide $2$. (b) Experimental scheme to characterize coupling constants over separation distance $d_i$, with $i=1,2,...,10$. (c) White light microscope image at the output facet of five S-S dimers with separation distances $16,17,18,19,20\ \mu$m. (d) Output images after laser (HeNe) excitation of waveguide $1$ [see red arrows in (b)].}
\label{sf5}
\end{figure}

Model (\ref{se2}) describes a homogeneous dimer for $\Delta\beta=0$. If waveguide $1$ is excited only, a periodic transfer of power is found: $$P_1(z)=P_0\cos^2(Cz)\ \ \text{and}\ \ P_2(z)=P_0\sin^2(Cz)\ ,$$ where $P_0=P_1(0)$ corresponds to the input power. The characterization method is based on this oscillatory dynamics [see Fig.~\ref{sf1}(b1)], but considering a very short evolution with less than one full cycle. The idea is to measure the transferred power from waveguide 1 to waveguide 2, when the coupling interaction is starting to occur. Fig.~\ref{sf5}(a) shows an example of this with three different coupling constants $C$ such that $C_1>C_2>C_3$. There, we observe that a smaller coupling constant implies a smaller amount of transferred power at an equal distance. Therefore, we experimentally extract coupling constants by fabricating dimer systems composed of a long waveguide 1 ($L_1=5$ cm) and a short waveguide 2 ($L_2=0.6$ cm) as sketched in Fig.~\ref{sf5}(b).
%

\noindent
We fabricate 10 dimers for each case (S-S, S-P, and P-P), for ten separation distances $d=16,17,...,25\ \mu$m. Fig.~\ref{sf5}(c) shows an example with 5 S-S dimers, where $d$ is growing to the right. Then, every dimer is excited by a HeNe laser excitation at waveguide 1 and output intensity images are taken with a CCD camera, as the examples shown in Fig.~\ref{sf5}(d). We analyze these images in a gray scale and extract the information about the intensity at waveguide 1 and the intensity at waveguide 2, as described in the main text. Then, we use power expressions to calculate the corresponding coupling constants as follows $$\frac{P_2(L_2,d)}{P_1(L_2,d)}=\tan^2(C L_2)\ \rightarrow\ C(d)=\frac{1}{L_2}\tan^{-1}\left(\sqrt{\frac{P_2(L_2,d)}{P_1(L_2,d)}}\right)\ .$$ In this way, we obtain coupling constants $C(d)$ as a function of separation distances $d$, and for every case under investigation, with the compiled information presented in Fig.3(a) (main text).

\section{Interferometric setup}

\begin{figure}[htbp]
\centering\includegraphics[width=14cm]{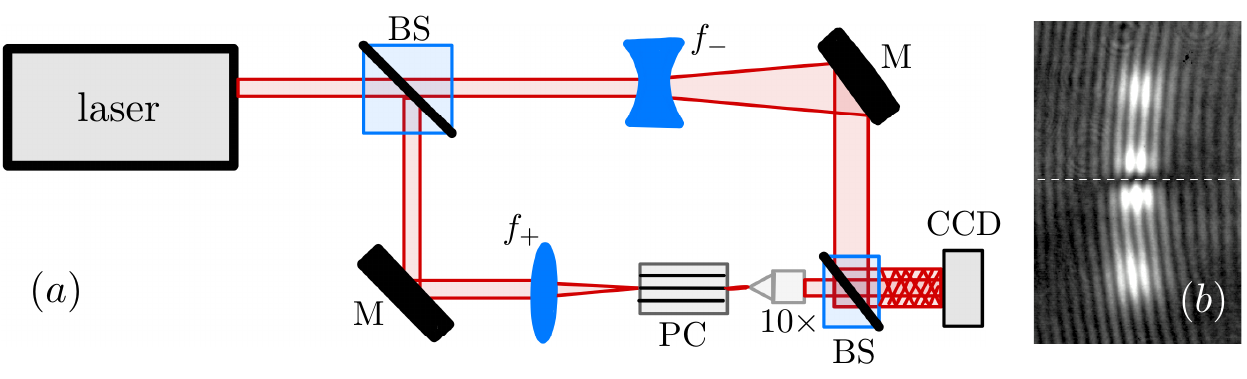}
\caption{Interferometer setup. (a) A laser beam at 632 nm is split into two beams with a beam splitter (BS). One beam is focused onto a photonic chip (PC) using a positive lens $f_+$. Output images are collected by a $10\times$ microscope objective onto a CCD camera. The upper part of the beam is expanded using a divergent lens $f_-$, which is then superposed with the output beam and observed using a CCD camera. (b) Interferogram of a wide gaussian beam with two S-mode waveguides (bottom and top) and a P-state at the center (see dashed line).}
\label{sf6}
\end{figure}

\noindent
The phase characterization presented in Fig.~3(d) (main text) is performed by a standard plane wave interferometer~\cite{sr6}, which is described in Fig.~\ref{sf6}(a). Figure~\ref{sf6}(b) shows an example, where upper and lower parts have a $\pi$ phase difference. A $\pi$ phase shift is obtained when a given interference maximum faces an interference minimum, as clearly shown in this figure.

{}